\documentclass[3p,times]{elsarticle}

\usepackage{ecrc}

\volume{00}

\firstpage{1}

\journalname{Nuclear Physics A}

\runauth{T. Lappi et al.}

\jid{nupha}

\jnltitlelogo{Nuclear Physics A}

\usepackage{graphicx}
\usepackage{amsmath,amssymb}

\usepackage{xcolor}
\definecolor{lcolor}{rgb}{0.5,0,0}
\definecolor{citcolor}{rgb}{0,0.3,0.0}
\usepackage[breaklinks,colorlinks,urlcolor=blue,citecolor=citcolor,linkcolor=lco
lor]{hyperref}
\usepackage{mciteplus}

\newcommand{\qs}{Q_\mathrm{s}}
\newcommand{\nc}{{N_\mathrm{c}}}
\newcommand{\as}{\alpha_{\mathrm{s}}}
\newcommand{\xt}{{\mathbf{x}_T}}
\newcommand{\nabt}{\boldsymbol{\nabla}_T}
\newcommand{\ud}{\, \mathrm{d}}
\newcommand{\tr}{\, \mathrm{Tr} \, }
\newcommand{\At}{\mathbf{A}_T}

\begin{document}

\begin{frontmatter}

\title{Structure of chromomagnetic fields in the glasma}

\author[tl1,tl2]{T. Lappi}
\address[tl1]{Department of Physics, %
 P.O. Box 35, 40014 University of Jyv\"askyl\"a, Finland}
\address[tl2]{Helsinki Institute of Physics, P.O. Box 64, 00014 University of Helsinki,
Finland}

\author[ad]{A. Dumitru}
\address[ad]{Department of Natural Sciences, Baruch College, New York, NY 10010, USA}
\author[yn]{Y. Nara}
\address[yn]{Akita International University, Yuwa, Akita-city 010-1292, Japan}




\begin{abstract}
The initial stage of a heavy ion collision is dominated by nonperturbatively strong chromoelectric and -magnetic fields. The spatial Wilson loop provides a gauge invariant observable to probe the dynamics of the longitudinal chromomagnetic field. We discuss recent results from a real time lattice calculation of the area-dependence of the expectation value of the spatial Wilson loop.
We show that at relatively early times after the collision, a universal scaling as a function of the area emerges at large distances for very different initial conditions, with a nontrivial critical exponent. A similar behavior has earlier been seen in calculations of the gluon transverse momentum spectrum, which becomes independent of the initial spectrum of gauge fields.
We also show the
distribution of eigenvalues of the spatial Wilson loop and the
fluctuations of its real and imaginary parts.
\end{abstract}

\begin{keyword}
Color glass condensate \sep Chromomagnetic field

\end{keyword}

\end{frontmatter}



\section{Introduction}\label{intro}
The initial stages of a heavy ion collision are dominated by nonperturbatively
strong gluonic fields. These fields are characterized by a momentum and length scale 
generated by nonlinear gluonic interactions, the saturation scale $\qs$. In the  
 ``Color Glass Condensate'' (CGC) effective theory description of QCD, the initial
``glasma''\cite{Lappi:2006fp} chromomagnetic and -electric fields are 
predominantly longitudinal~\cite{Lappi:2006fp,Kharzeev:2001ev,*Fries:2006pv}. 
Because of the  large ($\sim 1/\as$) occupation numbers of the gluonic states the 
glasma fields are essentially classical, and obey the 
Yang-Mills equations of motion. From these equations
 their time dependence can be solved
 either analytically in an expansion in the
field strength~\cite{Kovner:1995ja,*Blaizot:2008yb} or numerically on a
lattice~\cite{Krasnitz:1998ns,*Krasnitz:2001qu,*Krasnitz:2003jw,*Lappi:2003bi,Lappi:2007ku}.
Most earlier numerical studies have concentrated on the gluon spectrum, obtained
from Coulomb gauge-fixed field correlators. It would, however, be interesting to 
have an independent, manifestly gauge invariant way to study the dynamics of the 
softer field modes $k_T \lesssim \qs$. One possibility for such an observable,
first studied in this context in Ref.~\cite{Dumitru:2013koh}, is the spatial Wilson 
loop. We will here discuss results from a more recent calculation~\cite{Dumitru:2014nka}
that extends this work in many ways, with $\nc=3$ colors (instead of $\nc=2$
 in~\cite{Dumitru:2013koh}), measurements over a larger range of areas and most 
importantly a more systematical comparison of different initial conditions 
for the equations of motion, provided by different CGC parametrizations
of the two colliding nuclei. We stress that this work concerns 
the detailed structure of the boost invariant background fields and does not
include their unstable quantum fluctuations that eventually lead to an isotropization
of the system~\cite{Berges:2013eia,*Gelis:2013rba}.

\section{Classical Yang-Mills}

Before the collision the individual fields of projectile and target
are two dimensional pure gauges; in light cone gauge,
\begin{equation} \label{eq:alphai} 
\alpha^i_m(\xt) = \frac{i}{g} \, V_m(\xt)  \, \partial^i V_m^\dagger(\xt)
\end{equation}
where $m=1,\, 2$ labels the projectile and target, respectively. Here 
$V_m$ are light-like SU($\nc$) Wilson lines that describe the propagation 
of a lightlike probe through the color field; thus they can be related
to e.g. the DIS cross section. In the CGC they are stochastic variables
drawn from some probability distribution. 
We have in this calculation compared results from three well-motivated 
possibilities for this probability distribution. The first one 
is the MV model~\cite{McLerran:1994ni,*McLerran:1994ka,*McLerran:1994vd},
where 
the Wilson lines are obtained from a classical color
charge density $\rho$ as
\begin{equation} \label{eq:V_rho}
V(\xt) = \mathbb{P} \exp\left\{ i \int \ud x^-  
g^2 \frac{1}{ \nabt^2} \rho^a(\xt,x^-) \right\}, 
\end{equation}
where $\mathbb{P}$ denotes path-ordering in $x^-$.
In the MV model the densities $\rho$ are assumed to have a local Gaussian
distribution parametrized by a single dimensionful paramater $\mu$, related to 
the saturation scale as $\qs \sim g^2\mu$. The numerical implementation of the 
MV model in this context is described in detail in Ref.~\cite{Lappi:2007ku}.

At higher collision energies one needs to resum large corrections of order $\as \ln s$ 
from additional gluon brems\-strahlung. In the CGC framework this is achieved by the 
JIMWLK renormalization group equation, which describes the dependence
of the probability distribution of the Wilson lines  on $y\equiv \ln \sqrt{s}$.
The two other probability distributions studied in this work result from solving
the JIMWLK equation with either fixed or running  QCD coupling, with the MV model as 
an initial condition. For details on the numerical procedure used to do this we
refer the reader to Refs.~\cite{Blaizot:2002xy,*Rummukainen:2003ns,*Lappi:2012vw}.

\section{Results}

\begin{figure}
\begin{center}
\includegraphics*[width=0.45\textwidth]{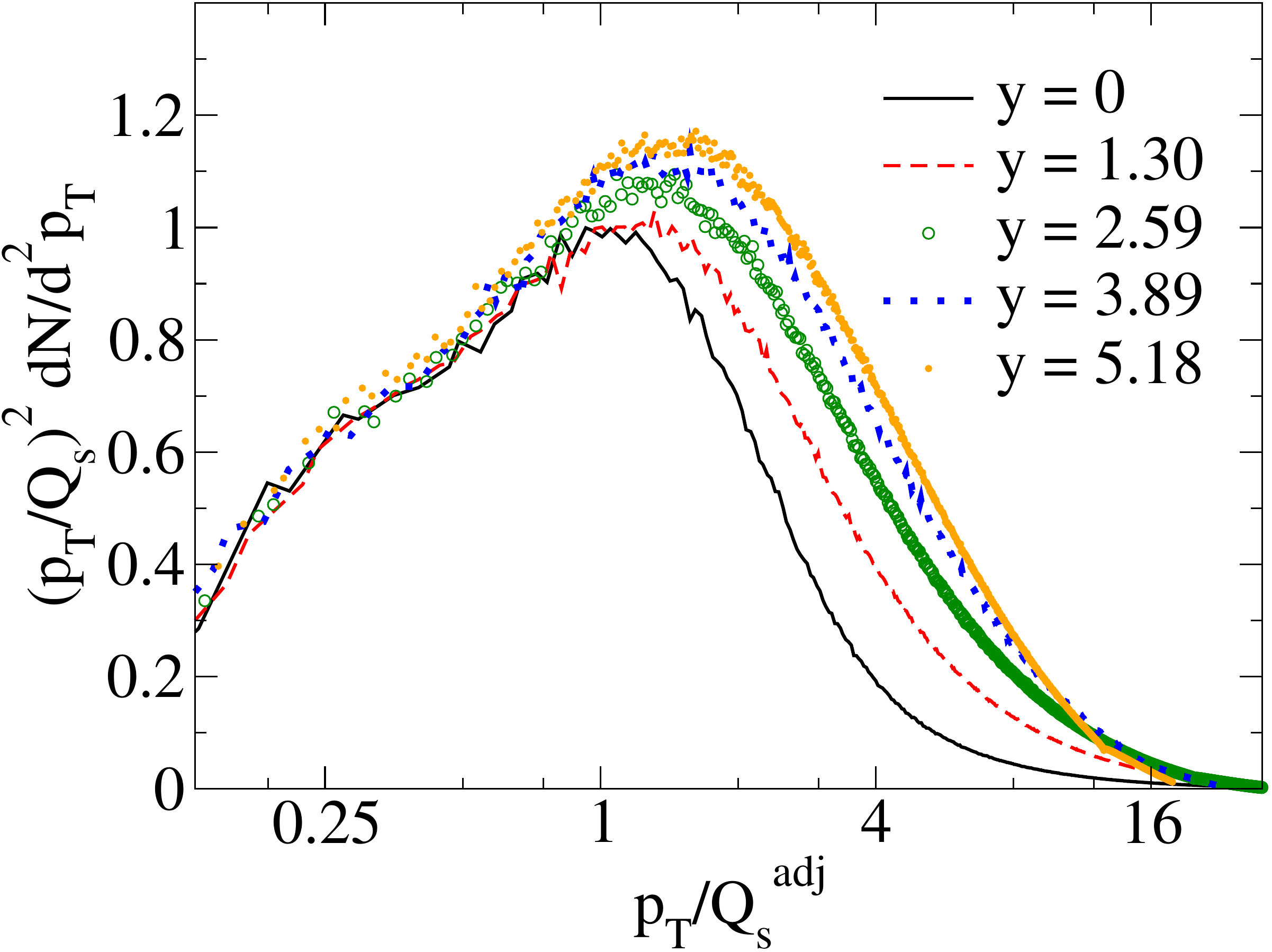}
\hfill
\includegraphics*[width=0.45\textwidth]{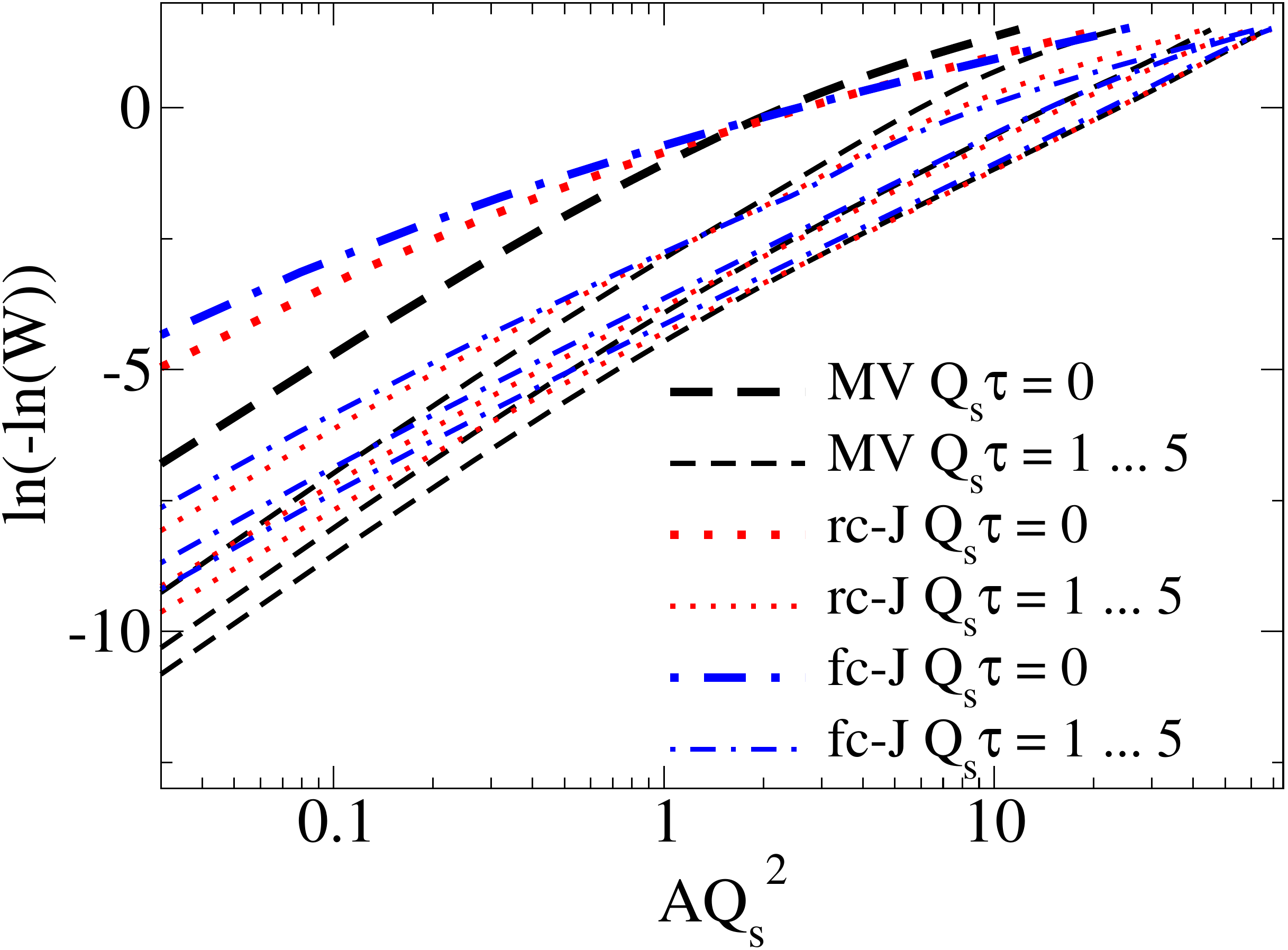}
\caption{Left: Gluon spectrum in the glasma in the symmetric (midrapidity) case,
expressed in terms of the scaling variable $p_T/\qs$.
 The initial condition at $y=0$ corresponds to 
the MV model, while after $y=\ln \sqrt{s} \approx 6$ units of JIMWLK evolution
(here at fixed coupling) the spectrum in the ultraviolet is different. 
Figure from Ref.~\cite{Lappi:2011ju}, with an error in the vertical axis
labeling corrected.
Right: Area dependence of the Wilson loop expectation value. The thick lines
are the initial condition and thinner ones correspond to
later times. Three different 
initial conditions are shown, corresponding to the MV model and fixed and running
coupling JIMWLK evolution, labeled ``fc-J'' and ``rc-J'' respectively.
}
\label{fig:spect}
\end{center}
\end{figure}
\begin{figure}
\begin{center}
\includegraphics*[width=0.45\textwidth]{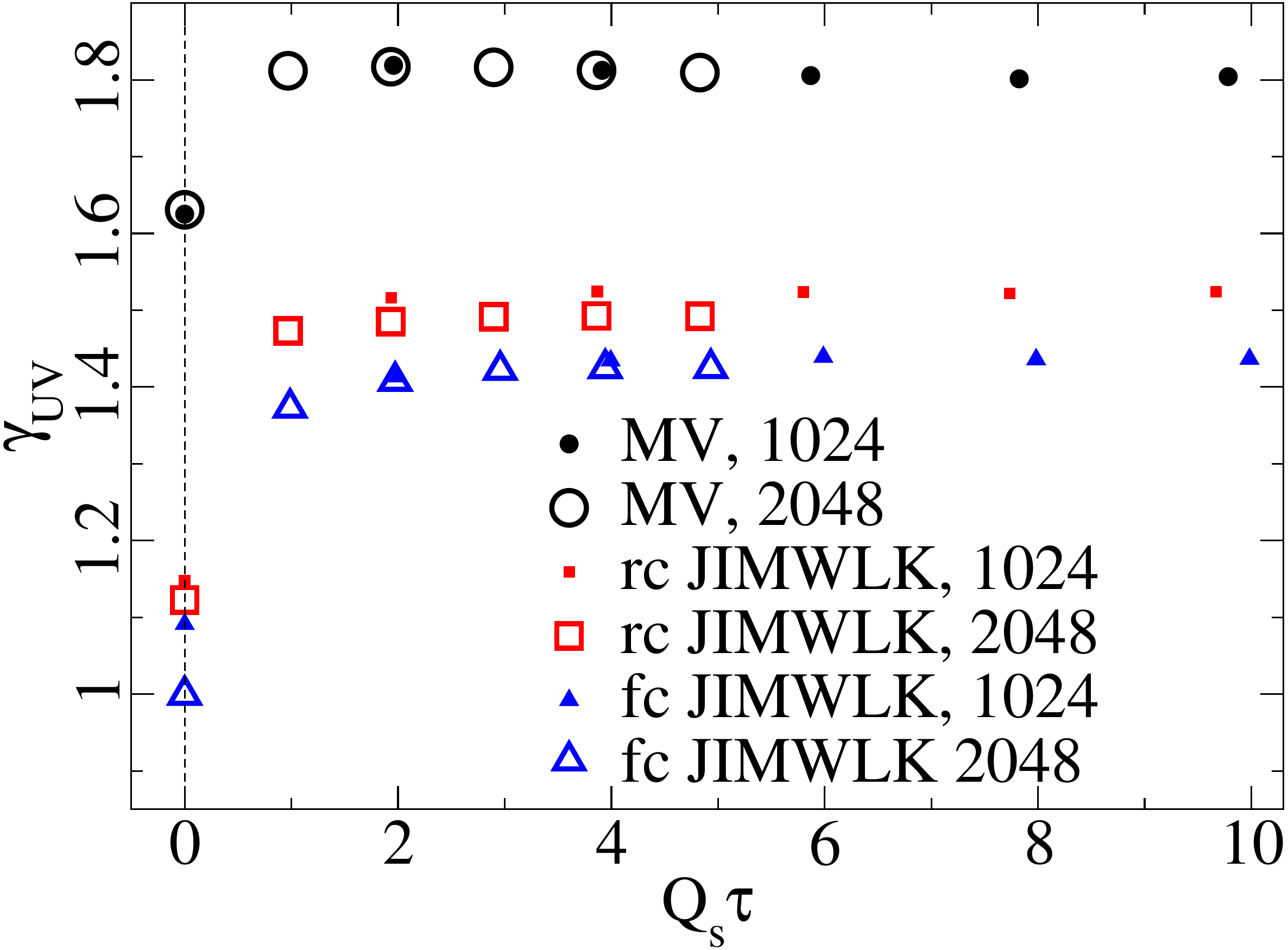}
\hfill
\includegraphics*[width=0.45\textwidth]{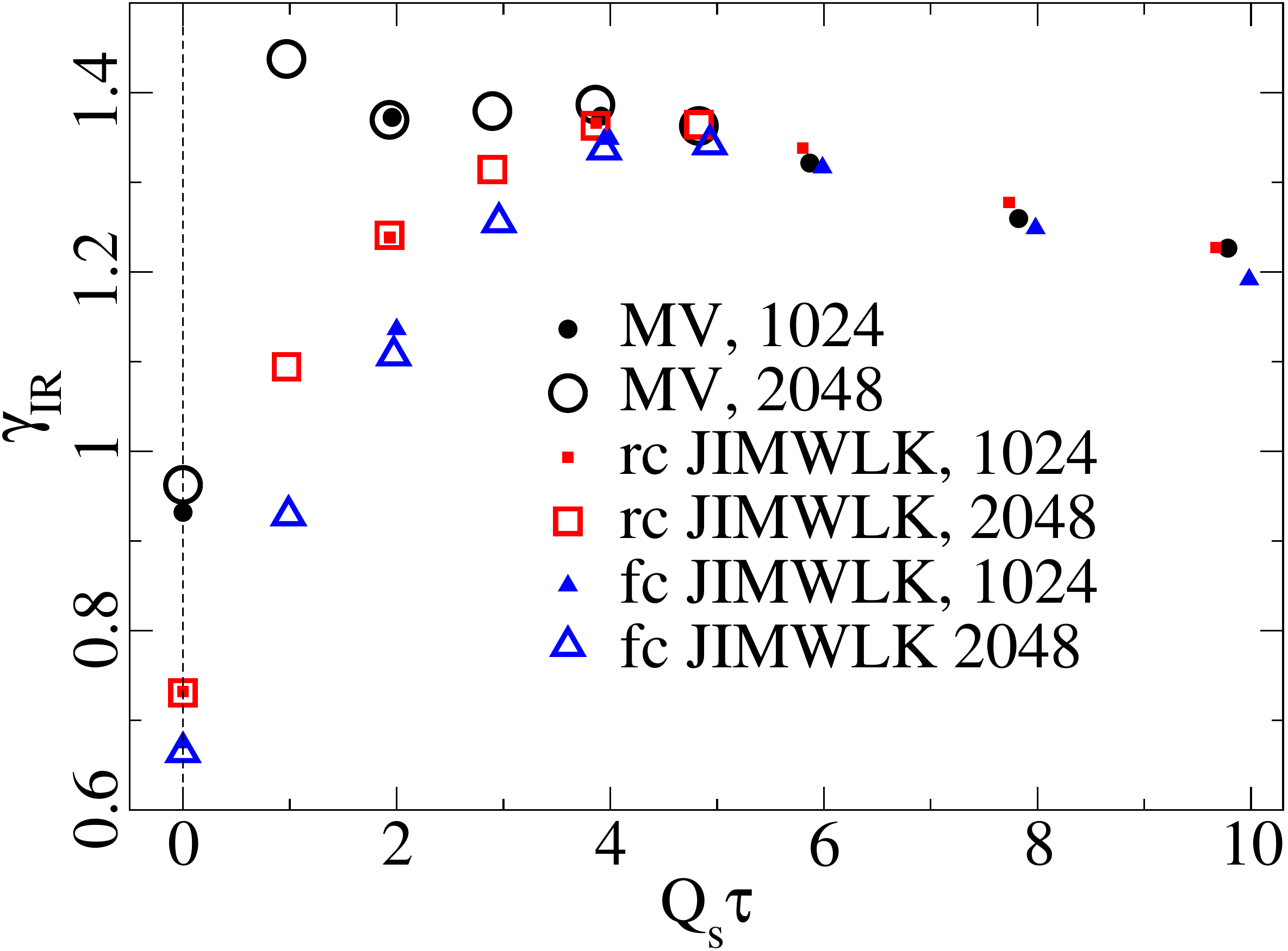}
\caption{The scaling exponents $\gamma$ for small area (left) and large area
(right) Wilson loops, as a function of time $\tau$. Comparison between MV, fixed and running coupling JIMWLK initial conditions and different lattice sizes.}
\label{fig:exps}
\end{center}
\end{figure}

Let us start by a reminder of earlier results for the gluon spectrum~\cite{Lappi:2011ju}, 
shown in Fig.\ref{fig:spect} (left). What was observed was the following: in the 
dilute $p_T\gtrsim \qs$ regime where final state interactions are relatively weak, the
spectrum depends on the initial conditions,  being harder with the JIMWLK distribution 
than the MV one, in agreement with expectations from $k_T$-factorization. In the infrared
part $p_T\lesssim \qs$, on the other hand, there is a remarkable universality,
both in the shape and the normalization, with 
$\ud N/\ud^2p_T \sim 1/p_T$ independently of the initial condition.

We then turn to the results for the spatial Wilson loop, which is defined as the
trace of a path ordered exponential of the gauge field around a closed
path of area $A$ in the transverse plane:
\begin{equation}
W(A) = \frac{1}{\nc} \left< \tr \mathbb{P} \exp\left\{i g 
\oint_{\partial A} \ud \xt \cdot \At \right\} \right>. 
\end{equation}
In Ref.~\cite{Dumitru:2014nka} we have found that the dependence of $W$ on the area $A$ of the loop 
is well described  by the functional form 
\begin{equation}\label{eq:fitform}
W(A) = \exp\left\{-(\sigma A )^\gamma \right\},
\end{equation}
with separate parameters $\sigma$ and $\gamma$ in the large (``IR'') and small (``UV'')
area regimes:
\begin{eqnarray}
\mathrm{IR:} && e^{0.5} < A \qs^2 < e^{5} \\
\mathrm{UV:} && e^{-3.5} < A \qs^2 < e^{-0.5}~.
\end{eqnarray}
Figure \ref{fig:spect} (right) shows the result for the Wilson loop expectation value,
for different initial conditions and different times $\tau$. For small area Wilson loops
the behavior parallels that of the gluonspectrum at high $p_T$: the differences in the 
initial condition manifest themselves in the area dependence, both
at the initial time $\tau=0$ and later. 
For large loops, the area dependence for different initial conditions is also different
at $\tau=0$. After a time $\tau \sim 3/\qs$, however, the simulations with different 
initial conditions
collapse onto a universal curve with both the slope
$\gamma$ and the normalization $\sigma$ independent of the initial condition. Interestingly, 
instead of the area law $\gamma=1$ that one might expect, the universal behavior is 
characterized by a nontrivial exponent $\gamma \approx 1.2$. The time dependence of the
scaling exponents is shown in Fig.~\ref{fig:exps}.

\section{Fluctuations of the Wilson lines}

\begin{figure}
\begin{center}
\includegraphics*[width=0.45\textwidth]{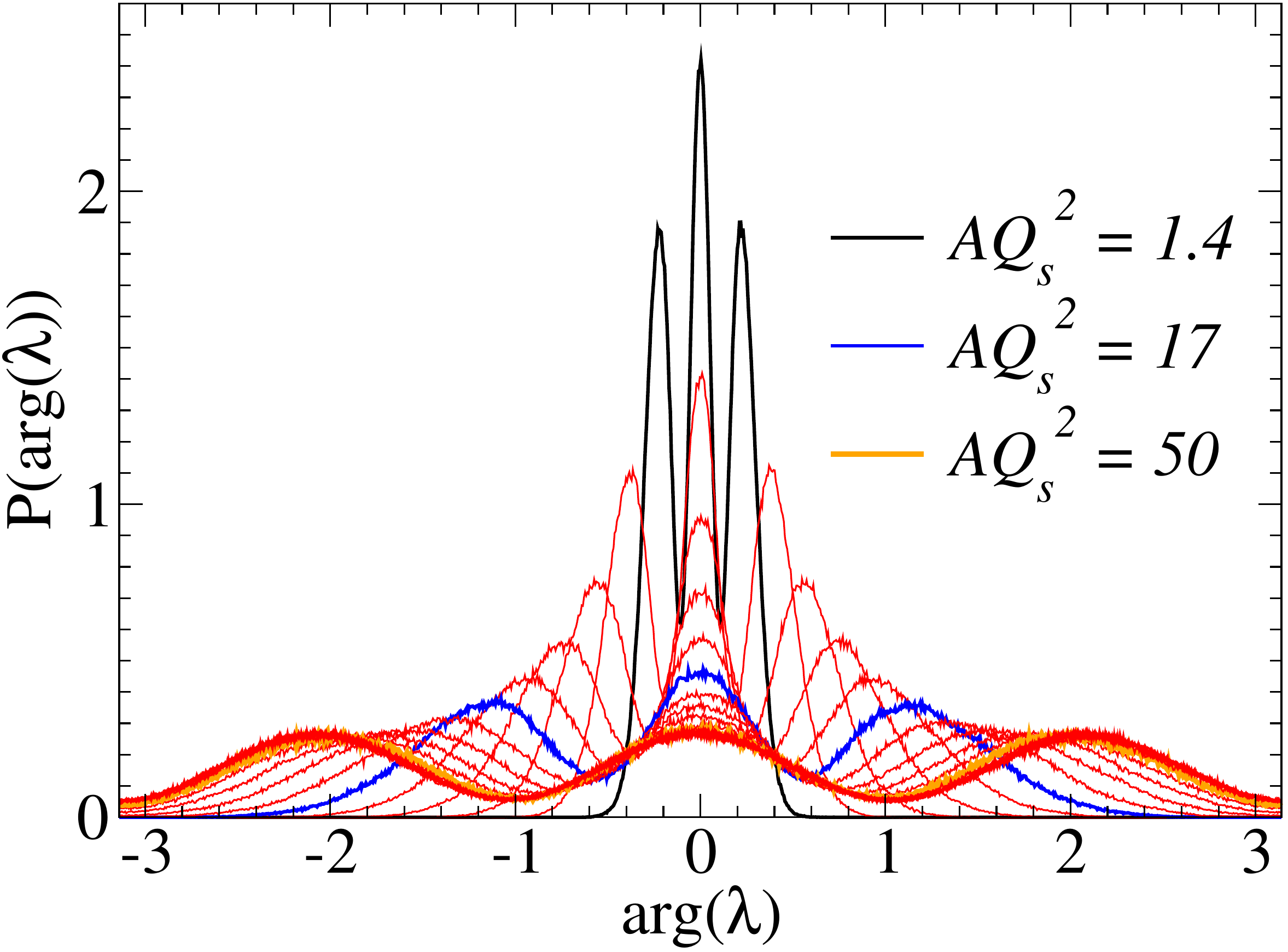}
\includegraphics*[width=0.45\textwidth]{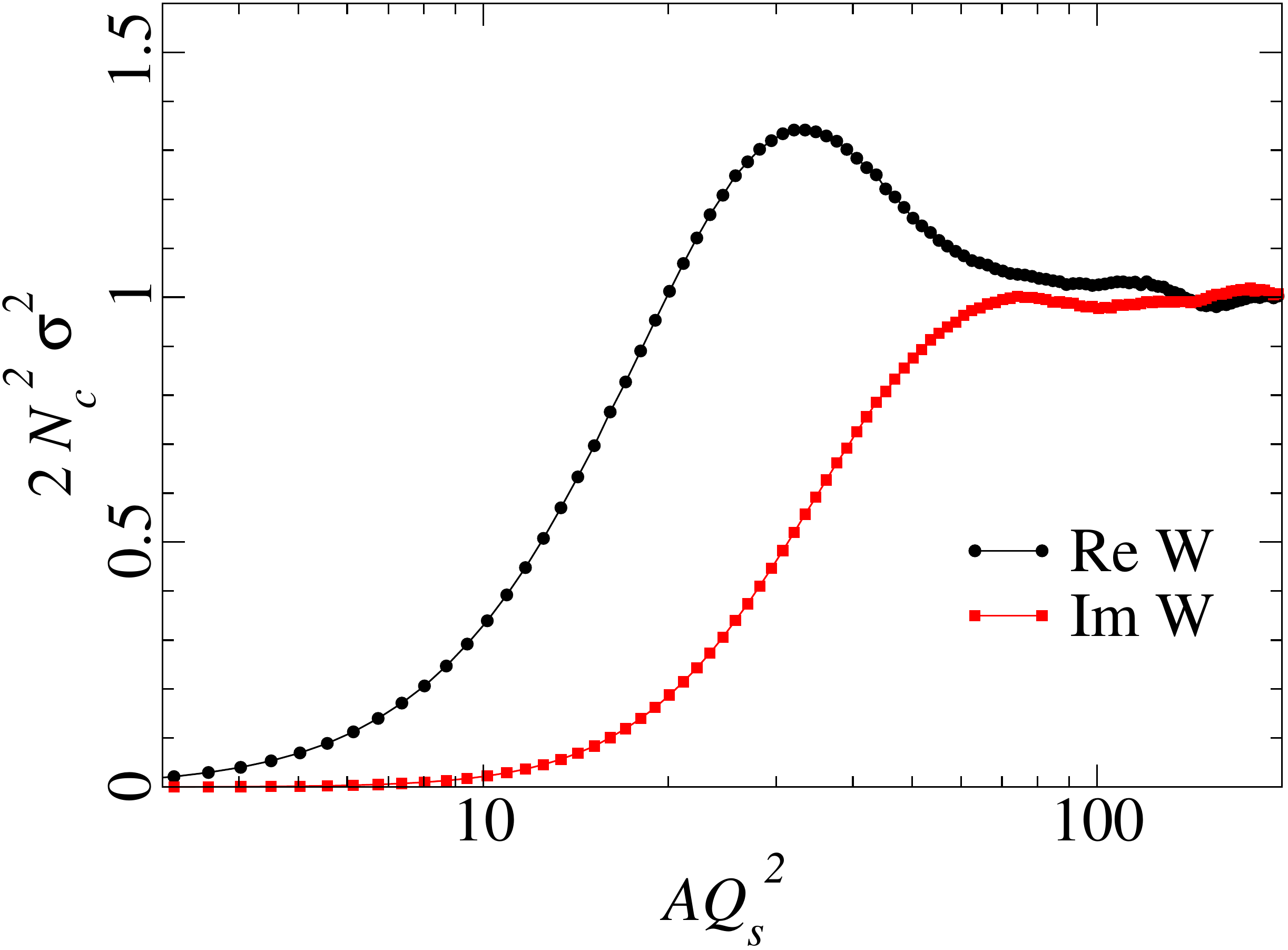}
\caption{Left: Probability distribution of the phase angle of the Wilson line eigenvalues
$\lambda$, for different loop areas.
 Right: variance of the real and imaginary part of the trace of the Wilson loop, 
as a function of the area of the loop. Simulations with running coupling JIMWLK 
initial conditions, at $\qs\tau=5$.
}
\label{fig:stats}
\end{center}
\end{figure}

As a further study beyond what was reported in Ref.~\cite{Dumitru:2014nka} 
we have recently looked at the statistical properties of the Wilson line
distribution in more detail. Figure \ref{fig:stats} (left) shows the probability
distribution of the eigenvalues of the Wilson lines $\lambda= e^{i \varphi}$.
One sees a progression from a three-peaked structure around $\varphi=0$ for small
areas (close to $W(A)=1$) to a distribution that is close to the expectation 
\begin{equation}
 P(\varphi) = \frac{1}{2\pi}\left(1 + \frac{2}{3} \cos(3\varphi) \right)
\end{equation}
from a completely random ensemble of SU(3) matrices for large areas.
Figure~\ref{fig:stats} (right) shows the variance of the real and imaginary parts of 
the trace of the Wilson loop. These start from a very small value for small 
areas, where the SU(3) matrix is constrained to be very close to the identity, 
and approach the random SU(3) matrix expectation for the variance
\begin{equation}
 \sigma^2 = \frac{1}{2\nc^2}
\end{equation}
for large areas.

\section*{Acknowledgements}
T.~L.\ has been supported by the Academy of Finland, projects 133005, 
267321 and 273464. This work was done using computing resources from
CSC -- IT Center for Science in Espoo, Finland.
A.~D.\ acknowledges support by the
DOE Office of Nuclear Physics through Grant No.\ DE-FG02-09ER41620 and
from The City University of New York through the PSC-CUNY Research
Award Program, grant 67119-00~45.

\bibliography{spires}
\bibliographystyle{h-physrev4mod2Mdense}

\end{document}